\documentclass[12pt]{article}
\usepackage{amssymb}

\newif\ifams\amsfalse
\amstrue \ifams
\fi
\newcommand{\bea}{\begin{eqnarray}}
\newcommand{\eea}{\end{eqnarray}}
\newcommand{\be}{\begin{equation}}
\newcommand{\ee}{\end{equation}}
\ifams

\else

\fi
\input{tcilatex}

\begin{document}

\title{%
\begin{flushright}
\begin{small}
 Alberta-Thy 02-02\\
 January 2002 \\
\end{small}
\end{flushright}
\vspace{1.cm} Rotational Dragging Effect on Statistical Thermodynamics
of (2+1)-dimensional Rotating de Sitter Space}
\author{Jeongwon Ho \\
{\small \vspace{- 3mm} Theoretical Physics Institute, Department of Physics,}
\\
{\small \vspace{- 3mm} University of Alberta, Edmonton, Canada T6G 2J1}\\
{\small \vspace{- 3mm} and}\\
{\small \vspace{- 3mm} Department of Physics, Kunsan National University,}\\
{\small \vspace{- 3mm} Kunsan, Korea 573-701}\\
{\small E-mail: \texttt{jwho@physicist.net}}}
\date{}
\maketitle

\begin{abstract}
Brief comments on a plausible holographic relationship between the opposite
rotational dragging effect of a (2+1)-dimensional rotating de Sitter space
and the non-unitarity of a boundary conformal field theory are given. In
addition to the comments, we study how the opposite rotational dragging
effect affects the statistical-mechanical quantities in the rotating de
Sitter space in comparison with a BTZ black hole.
\end{abstract}

\newpage

Gibbons and Hawking \cite{gibbons77a} have shown that an inertial observer
in de Sitter (dS) space detects thermal radiation at temperature
$\kappa/2\pi $ coming from the cosmological horizon and the
entropy of dS space satisfies the Bekenstein-Hawking area law given by
\begin{equation}
S=\frac{A}{4G}, 
\label{bhlaw}
\end{equation}
where $\kappa $, $A$ are the surface gravity and the area of the
cosmological horizon, respectively. This result extended the validity
of classical and semiclassical thermodynamic properties of black holes to
cosmological spaces. Recently, Gibbons and Hawking's work has been
reviving at a microscopic level; guided by analogies to the anti-de Sitter
(AdS)/conformal field theory (CFT) correspondence \cite{adscft}, one has
obtained the entropy of dS space (\ref{bhlaw}) by naively using the Cardy
formula under the assumption of a boundary CFT, which is dual to quantum
gravity on
dS space \cite{dscft}.

In spite of many fruitful observations of the conjectured dS/CFT
correspondence in analogy to the AdS/CFT, there are strict
and important differences between the two correspondences. One is
the appearance of complex conformal weights in the dS/CFT, which reflects
non-unitarity of the dual CFT \cite{strominger0106}.
It has been shown that in the case of (2+1)-dimensional rotating dS (RdS)
spaces non-unitarity of the dual CFT may be related to rotational motion
of the bulk space \cite{muin98}\cite{bala0110}. Strictly speaking,
imaginary part of eigenvalues of conformal generators $L_{0}$ and
$\overline{L}_0$ in stationary coordinates is given by the angular momentum
of the RdS space. Thus, taking the limit of zero angular momentum,
the unitarity of the dual CFT should be recovered. While on the other,
the boundary CFT, which is dual to a BTZ black hole \cite{btz}, has real
eigenvalues for $L_{0}$ and $\overline{L}_0$ even for non-zero angular
momentum \cite{strominger98}. Therefore, it is suspected that somehow the
rotational effect of (2+1)-dimensional RdS space holographically triggers non-unitarity in
the dual CFT.

What is the geometrical difference between RdS spaces and rotating AdS black
holes? The most distinct geometrical aspect of spacetimes around rotating
objects is the rotational dragging effect on inertial frames around the
object. Inside the cosmological horizon of RdS spaces, in contrast to
spacetimes outside rotating AdS black holes (and asymptotically flat
rotating black holes), the rotational dragging effect is exerting in the
opposite direction of space rotation.\footnote{A detailed comment is given
in the paragraph between Eq.(\ref{minte}) and Eq.(\ref{zamo}).} Thus, it is
likely that \textit{the non-unitarity of the boundary
CFT, which is dual to (2+1)-dimensional RdS space, may be holographically
related to the opposite rotational dragging effect (inside the cosmological
horizon) of the bulk space}.

An objection may be raised to the suggestion. Since the conjectured
dual CFT lives in the future and/or past infinity outside the cosmological
horizon, the rotational dragging effect inside the horizon cannot affect
the boundary CFT. A plausible answer to this objection is directly related
to one of the main issues of holographic principle, non-locality: according
to the holographic principle, an event in a bulk spacetime, which is causally
disconnected to the boundary from the bulk viewpoint, has to be encoded in
the boundary dual theory in a non-local way. The boundary variables in which
such events are encoded are called `precursors' \cite{precursor}.
Thus, it seems that the opposite rotational dragging effect inside the
cosmological horizon may be encoded in the boundary theory in a non-local way
and triggers the non-unitarity of the boundary theory. We leave the proof of
the idea to a future work.

Adopting a strong interpretation of non-locality, thermodynamics of the
cosmological horizon, which has been formulated inside the horizon
\cite{gibbons77a}, may be reconstructed outside the horizon by
requiring some appropriate prescriptions. In fact, an analogue of the
Gibbs-Duhem relation was obtained outside the cosmological horizon in
\cite{mann0111}. This statement is closely related to the concept of
entanglement entropy, which is a strong candidate for the
statistical-mechanical origin of the black hole entropy. (As a matter of
fact, non-locality is a consequence of quantum entanglement.) For instance,
if a bipartite system is a pure state, its two subsystems (e.g. inside and
outside the horizon) have the same entanglement entropy \cite{muko97}.

In this sense, it seems reasonable to study how the opposite rotational
dragging effect affects the statistical-mechanical quantities in a
(2+1)-dimensional RdS space comparing with the case of a BTZ black hole
\cite{ho99}. For that purpose, we shall use the brick wall model
\cite{thooft85}, which is strongly supported by the concept of entanglement
entropy. The statistical-mechanical entropy of quantum fields in a
(2+1)-dimensional non-rotating de Sitter space has been calculated in
\cite{wtkim98}.

The (2+1)-dimensional RdS space is described by the metric
\begin{equation}
ds^{2}=-N^{2}dt^{2}+N^{-2}dr^{2}+r^{2}(d\varphi +N^{\varphi }dt)^{2},
\label{metric}
\end{equation}
where
\begin{equation}
N^{2}=M-\frac{r^{2}}{l^{2}}+\frac{J^{2}}{4r^{2}}=\frac{
(r^{2}-r_{+}^{2})(r^{2}+r_{-}^{2})}{r^{2}l^{2}},~~N^{\varphi }=-\frac{J}{
2r^{2}},
\label{nftn}
\end{equation}
and $r_{+}=l\sqrt{M/2}(1+(1+J^{2}/M^{2}l^{2})^{1/2})^{1/2}$ denotes the
cosmological horizon. We also introduced a positive parameter $r_{-}$ defined
by $r_{-}^{2}\equiv -(Ml^{2}/2)(1-(1+J^{2}/M^{2}l^{2})^{1/2})$. In terms of
$r_{\pm }$, the mass and angular momentum \cite{bala0110} can be rewritten by
$M=(r_{+}^{2}-r_{-}^{2})/l^{2}$ and $J=2r_{+}r_{-}/l$, respectively.
Note that an ergoregion exists inside the horizon $r_{erg}=l\sqrt{M}<r_{+}$.

In the original brick wall model, a field is confined to a shell region between
inner and outer walls and satisfies the periodic boundary condition
\begin{equation}
\Phi (r_{+}-h)=\Phi (L).
\label{BC}
\end{equation}
Here, $h$ is the brick wall cutoff, and $r_{+}-h$ the `outer' wall and
$L$ $(<r_{+}-h)$ the `inner' wall. In the dS space the volume inside the
horizon is finite, so there are no infrared divergences. Thus, the infrared
cutoff $L$ is not necessary in our calculation \cite{wtkim98}. We consider
a massless scalar field satisfying the Klein-Gordon equation $\square \Phi
=0 $. The mode solutions of the Klein-Gordon equation is given by
$\Phi (t,r,\varphi )=R_{Em}(r)e^{-iEt+im\varphi }$. The
radial part $R_{Em}(r)$ satisfies 
\begin{equation}
\frac{1}{\sqrt{-g}}\frac{d}{dr}\left(
\sqrt{-g}g^{rr}\frac{dR_{Em}}{dr} \right) +g^{rr}k^{2}(r;E,m)R_{Em}=0,
\label{REQ}
\end{equation}
where
\begin{equation}
k^{2}(r;m,E)\equiv g_{rr}\left(
\frac{g_{\varphi \varphi }}{-D}E^{2}+2\frac{
g_{t\varphi }}{-D}mE+\frac{g_{tt}}{-D}m^{2}\right) ,
\label{RWN}
\end{equation}
and $D\equiv g_{tt}g_{\varphi \varphi }-g_{t\varphi }^{2}$. In the WKB
approximation, the radial quantum number $n(E,m)$ with energy $E$ and
angular momentum $m$ is given by the radial wave number $k(r;E,m)$ in
(\ref{RWN})
\begin{equation}
\pi n(E,m)=\int_{0}^{r_{+}-h}dr^{\prime }k^{\prime }(r;E,m),
\label{NMS}
\end{equation}
where $^{\prime }k^{\prime }(r;E,m)$ is set to be zero if $k^{2}(r;E,m)$
becomes negative for given $(E,m)$ \cite{thooft85}. Then, from the definition
of the density function $g(E,m)=\partial n(E,m)/\partial E$, $g(E,m)dE$
represents the number of single-particle states whose energy lies between $E$
and $E+dE$, and whose angular momentum is $m$.

The free energy is obtained by using the single-particle spectrum.
Due to the presence of ergoregion, scalar fields near the rotating
horizon have superradiant (SR) mode solutions. In this case,
the free energy of the system can be decomposed into the non-superradiant (NSR)
modes part $F_{\mathrm{NSR}}$ and the SR modes part $F_{\mathrm{SR}}$,
$F=F_{\mathrm{NSR}}+F_{\mathrm{SR}}$ \cite{ho99}. Following the formulation
given in \cite{ho99}, we write the NSR and SR parts of the free energy as
following;
\begin{eqnarray}
F_{\mathrm{NSR}} &=&\sum_{\mathrm{NSR}}\int dEg(E,m)\ln \left[ 1-e^{-\beta
(E-\Omega _{H}m)}\right] ,  \label{FNSR} \\
F_{\mathrm{SR}} &=&\sum_{\mathrm{SR}}\int dEg(E,m)\ln \left[ 1-e^{\beta
(E-\Omega _{H}m)}\right] ,  \label{FSR}
\end{eqnarray}
where $\Omega _{H}$ is the angular speed of the horizon. Here, we assume
that the scalar field is in a thermal equilibrium state at temperature
$\beta ^{-1}$ and all states with $E-\Omega _{H}m<0$ belong to the SR
modes and others to the NSR modes.

Performing the integration, we obtain the NSR part of the free energy given by
\begin{equation}
F_{\mathrm{NSR}}=F_{\mathrm{NSR}}^{(m>0)}+F_{\mathrm{NSR}}^{(m<0)},
\label{NSR2}
\end{equation}
where
\begin{eqnarray}
F_{\mathrm{NSR}}^{(m>0)} &=&-\frac{\Gamma (3)\zeta (3)}{\pi \beta ^{3}}
\int_{0}^{r_{+}-h}dr\left( \frac{g_{rr}g_{\varphi \varphi }}{-D}\right)
^{1/2}K_{\mathrm{NSR}}^{(m>0)}  \label{NSRP} \\
F_{\mathrm{NSR}}^{(m<0)} &=&-\frac{1}{\pi \beta }\int_{r_{erg}}^{r_{+}-h}dr
\left( \frac{g_{rr}g_{tt}}{-D}\right) ^{1/2}\int_{0}^{\infty }dmm\ln \left(
1-e^{-\beta \Omega _{H}m}\right)  \nonumber \\
&&-\frac{\Gamma (3)\zeta (3)}{\pi \beta ^{3}}\int_{r_{erg}}^{r_{+}-h}dr
\left( \frac{g_{rr}g_{\varphi \varphi }}{-D}\right)^{1/2}
K_{(m<0)}^{\mathrm{NSR}},  \label{NSRN}
\end{eqnarray}
where the functions $K_{\mathrm{NSR}}$ are defined by
\begin{eqnarray}
K_{\mathrm{NSR}}^{(m>0)} &=&\frac{g_{t\varphi }+\Omega _{H}g_{\varphi
\varphi }}{-2\tilde{g}}+\frac{-D}{2(-\tilde{g})^{3/2}g_{\varphi \varphi
}^{1/2}}\left( \sin ^{-1}\frac{g_{t\varphi }
+\Omega _{H}g_{\varphi \varphi }}{(-D)^{1/2}}+\frac{\pi }{2}\right) ,
\label{ftnF1} \\
K_{\mathrm{NSR}}^{(m<0)} &=&\frac{(g_{t\varphi }+\Omega _{H}g_{\varphi
\varphi })(\Omega _{H}-(g_{tt}/g_{\varphi \varphi })^{1/2})}{
2\tilde{g}\Omega _{H}}
+\left( \frac{g_{tt}}{g_{\varphi \varphi }}\right)^{1/2}
\frac{1}{2\Omega _{H}^{2}}  \nonumber \\
&&+\frac{-D}{2(-\tilde{g})^{3/2}g_{\varphi \varphi }^{1/2}}\left( \sin ^{-1}
\frac{g_{t\varphi }+\Omega _{H}g_{\varphi \varphi }}{-(-D)^{1/2}}-\sin ^{-1}
\frac{g_{tt}+\Omega _{H}g_{t\varphi }}{(-D)^{1/2}\Omega _{H}}\right) ,
\end{eqnarray}
where $\tilde{g}\equiv g_{\varphi \varphi }\Omega _{H}^{2}
+2g_{t\varphi}\Omega _{H}+g_{tt}$. The superradiant part of the free energy
becomes
\begin{eqnarray}
F_{\mathrm{SR}} &=&+\frac{1}{\pi \beta }\int_{r_{erg}}^{r_{+}-h}dr\left(
\frac{g_{rr}g_{tt}}{-D}\right) ^{1/2}\int_{0}^{\infty }dmm\ln \left(
1-e^{-\beta \Omega _{H}m}\right)  \label{SR2} \\
&&-\frac{\Gamma (3)\zeta (3)}{\pi \beta ^{3}}\int_{r_{erg}}^{r_{+}-h}dr
\left( \frac{g_{rr}g_{\varphi \varphi }}{-D}\right) ^{1/2}K_{\mathrm{SR}},
\end{eqnarray}
where
\begin{equation}
K_{\mathrm{SR}}=\frac{\Omega _{H}g_{t\varphi }+g_{tt}}{-2\tilde{g}\Omega
_{H}^{2}}\left( \frac{g_{tt}}{g_{\varphi \varphi }}\right) ^{1/2}+\frac{-D}{
2(-\tilde{g})^{3/2}g_{\varphi \varphi }^{1/2}}\left( \sin ^{-1}\frac{\Omega
_{H}g_{t\varphi }+g_{tt}}{(-D)^{1/2}\Omega _{H}}+\frac{\pi }{2}\right) .
\label{ftnF2}
\end{equation}
Then, neglecting the terms including the integration range from $0$ to
$r_{erg}$, the total free energy is given by a simple form
\begin{eqnarray}
F &=&F_{\mathrm{NSR}}+F_{\mathrm{SR}} \nonumber  \\
&=&-\frac{\zeta (3)}{\beta ^{3}}\int_{r_{erg}}^{r_{+}-h}dr\frac{%
(-g_{rr}D)^{1/2}}{(-\tilde{g})^{3/2}}. \label{finalF}
\end{eqnarray}

Now, we are ready to derive thermodynamic quantities of the scalar
field in thermal equilibrium with the RdS space. At first, the entropy
of the scalar field is obtained by
\begin{eqnarray}
S &=&\beta ^{2}\frac{\partial F}{\partial \beta }{\bigg|}_{\beta =\beta _{H}}
\nonumber \\
&=&\left( \frac{3\zeta (3)}{16\pi ^{3}\epsilon }\right) (2\cdot 2\pi r_{+}),
\label{entropy}
\end{eqnarray}
where $\beta _{H}=2\pi l^{2}r_{+}/(r_{+}^{2}+r_{-}^{2})$ is the inverse
Hawking temperature of the RdS space and $\epsilon $ is the proper
distance of the brick wall from the horizon given by 
\begin{equation}
\epsilon =\int_{r_{+}-h}^{r_{+}}drN^{-1}(r)\approx \left( \frac{2l^{2}r_{+}}{%
r_{+}^{2}+r_{-}^{2}}\right) ^{1/2}\sqrt{h}.  \label{pplength}
\end{equation}%
Thus, choosing the cutoff of the proper distance as
\begin{equation}
\epsilon =3\zeta (3)/(16\pi ^{3}),  \label{cutoff}
\end{equation}%
the entropy given by (\ref{entropy}) satisfies the Bekenstein-Hawking area
law. Note that the brick wall cutoff in (\ref{cutoff}) is equal to the
cutoff of the BTZ black hole \cite{ho99}. In addition, as in the case of BTZ
black hole, the SR contribution makes the cutoff twice as
much as that of non-rotating (2+1)-dimensional dS space.
(The free energy of SR modes (\ref{FSR}) is equal to that of NSR modes
(\ref{FNSR}) up to the leading term. See \cite{ho99}.)

Using the cutoff value given by (\ref{cutoff}), the angular momentum $J_{m}$
and internal energy $U_{m}$ of the scalar field are obtained as
\begin{eqnarray}
\label{mang} 
J_{m} &=&-\frac{\partial F}{\partial \Omega }{\bigg|}_{\beta =\beta
_{H},\Omega =\Omega _{H}}=-J,  \\
\label{minte}
U_{m} &=&\frac{\partial }{\partial \beta }(\beta F){\bigg|}_{\beta =\beta
_{H}}+\Omega _{H}J_{m}=\frac{4}{3}M+\frac{1}{3}\Omega _{H}J.
\end{eqnarray}
In fact, these thermal wall contributions, which is interpreted as the
backreaction of background space, lead a fatal flaw to the brick wall model
in which the background geometry is fixed \cite{susskind94}. This problem,
however, can be resolved by taking the Boulware state as a ground state in
the model \cite{mukohyama98}. In this `topped-up' Boulware description of
the brick wall model, the thermal contributions given by (\ref{mang}) and
(\ref{minte}) are to be canceled by the Boulware energy up to an appropriate
order of mass and angular momentum. However, since our purpose in this
report is to examine the thermal properties of quantum fields, the
regularization scheme is not considered here.

First of all, the angular momentum in Eq.(\ref{mang}) is the negative value
of the dS space angular momentum. Note that the angular momentum of quantum
fields on the BTZ black hole is the positive value of the BTZ black hole
angular momentum, $J_{m}^{BTZ}=J^{BTZ}$ \cite{ho99}. The negative sign in
(\ref{mang}) can be interpreted as following; inside RdS spaces the angular
speed $\Omega =J/(2r^{2})$ \textit{decreases} as approaching to the
cosmological horizon $r\rightarrow r_{+}$ $(r<r_{+})$. This means that
the rotational dragging effect of the space reduces the angular speed of a test
particle as approaching to the horizon, and entering the
ergoregion $r_{erg}<r<r_{+}$ the particle cannot rotate with angular speed
\textit{greater} than the angular speed of the horizon. In other words,
the direction of the dragging effect is opposite to the direction of the
horizon angular velocity. Therefore, the minus sign in (\ref{mang}) indicates
the opposite rotational dragging effect of the RdS space.

The opposite rotational dragging effect also arises in the internal energy
given by Eq.(\ref{minte}). The internal energy is enhanced with the positive
term $+\Omega _{H}J/3$, while in the case of BTZ (and Kerr) black hole,
rotation reduces internal energy, $U_{m}^{BTZ}=4M^{BTZ}/3-\Omega
_{H}J^{BTZ}/3$ \cite{ho99}. This argument looks strange, because the
rotational energy $\Omega _{H}J_{m}$ in (\ref{minte}) is negative in the RdS
space and positive in the BTZ black hole. In order to interpret this result,
consider the internal energy of the system with respect to a
zero-angular-momentum-observer (ZAMO) $U_{m}^{ZAMO}$. Since the rotation of
the dS space enlarges the radius of the cosmological horizon $r_{+}$
(thus, increases entropy), the ZAMO's internal energy receives a positive
rotational contribution
\begin{equation}
U_{m}^{ZAMO}=\frac{\partial }{\partial \beta }(\beta F){\bigg|}_{\beta
=\beta _{H}}=\frac{2}{3}\frac{S}{\beta _{H}}=\frac{4}{3}M+\frac{4}{3}\Omega
_{H}J. 
\label{zamo}
\end{equation}
Note that Eq.(\ref{zamo}) (and Eq.(\ref{minte})) is equivalent to the
Gibbs-Duhem relation and the factor $2/3$ of the term $S/\beta _{H}$ is
related to the equation of state $\rho _{m}\simeq 2P_{m}$, where
$\rho _{m}$ and $P_{m}$ denote the energy density and pressure, respectively.
Then, turning to Eq.(\ref
{minte}), the negative rotational energy
$\Omega _{H}J_{m}=-\Omega _{H}J$ does not eliminate the whole amount of
the positive rotational effect in the internal energy of ZAMO,
$+4\Omega _{H}J/3$. In the case of BTZ (Kerr) black hole, the rotational
effect decreases entropy and the positive rotational energy
$\Omega _{H}^{BTZ}J_{m}^{BTZ}=+\Omega _{H}^{BTZ}J^{BTZ}$ compensates partially
the negative rotational effect in the internal energy of ZAMO,
$-4\Omega_{H}^{BTZ}J^{BTZ}/3$.

In summary, we have studied how the opposite rotational dragging effect of
the RdS space affects thermodynamic quantities of quantum fields on the
RdS space. The effect arising in the internal energy and angular
momentum of thermal excitations is opposite to the case of BTZ black hole.
If the conjectured dS/CFT is true, these statistical-mechanical
properties inside the cosmological horizon should be encoded in the boundary
theory. Our expectation is that these properties would be closely related
to the non-unitarity of the boundary CFT. As a byproduct we have shown that
even though the appearance of SR modes is due to the presence of ergoregion,
the direction of rotational dragging effect does not affect to the SR
contribution to the brick wall cutoff, i.e. the SR contribution
makes the cutoff twice as much as that of non-rotating dS space as in the
case of BTZ black hole.

\vspace{0.2in} \textbf{Acknowledgments:} This work was supported in part
by the Natural Science and Engineering Research Council of Canada and by
Korea Science and Engineering Foundation under Grant No. 1999-2-112-003-5.
The author would like to thank S.P. Kim, C.H. Lee, S. Mukohyama, D.N. Page,
and M.I. Park for many useful discussions.

\end{document}